\documentclass[sigconf,screen,nonacm]{acmart}

\setcopyright{rightsretained}
\acmYear{2026}

\usepackage{booktabs,array,tabularx,graphicx,xcolor,enumitem,tikz,pgfplots}
\pgfplotsset{compat=1.18}
\usepgfplotslibrary{statistics}
\usetikzlibrary{arrows.meta,positioning,fit,calc}
\setlist[itemize]{leftmargin=*,topsep=1pt,itemsep=0pt,parsep=0pt,partopsep=0pt}
\setlist[enumerate]{leftmargin=*,topsep=1pt,itemsep=0pt,parsep=0pt,partopsep=0pt}

\definecolor{accentblue}{HTML}{006D9C}
\definecolor{accentteal}{HTML}{0F766E}
\definecolor{accentorange}{HTML}{D55E00}
\definecolor{accentpurple}{HTML}{7556A6}
\definecolor{slate}{HTML}{64748B}
\definecolor{heat4}{HTML}{BBF7D0}
\definecolor{heat5}{HTML}{86EFAC}
\definecolor{heat6}{HTML}{4ADE80}

\newcommand{\code}[1]{\texttt{#1}}
\newcommand{\spdf}{\ensuremath{S_{\mathrm{pdf}}}}

\hypersetup{urlcolor=black}

\begin{document}

\title{Engine-Transfer-Bench: An Evidence-Based Benchmark for Document
Compilation Engine Selection}

\author{Prajwal S. Venkateshmurthy}
\orcid{0009-0004-8830-0725}
\affiliation{%
  \institution{Independent Researcher}
  \city{San Jose}
  \state{CA}
  \country{USA}
}
\email{prajwal@prajwal.me}

\begin{abstract}
There is no shared framework for selecting among document compilation engines
(pdfLaTeX, XeLaTeX, LuaLaTeX, Tectonic, Typst, pandoc PDF backends).
We present Engine-Transfer-Bench (ETB): 1{,}784 open documents, four tasks
(reliability, latency, text consistency, failure modes), a pinned harness,
and host-tagged multi-OS results, to our knowledge the first cross-platform
empirical study of document compilation engines under identical seeds and
protocol. On GitHub Actions ($N{=}4{,}211$ compiles per host on macOS,
Ubuntu, and Windows), Tectonic's success rate is stable within 0.9\,pp
(96.3--97.2\%), while classic TeX~Live-style engines swing 12--20\,pp with
distribution policy (Ubuntu apt vs.\ MiKTeX auto-install vs.\ macOS
BasicTeX). On 702 portable LaTeX documents every engine tested
succeeds, so latency decides. Failures concentrate in 107 engine-specific
templates and are architectural (fonts, layout, assets) rather than missing
packages on a provisioned host. We release ETB, the ETB-Porta recommender,
and a public cross-OS harness as shared infrastructure.
\end{abstract}



\maketitle

\section{Introduction}\label{sec:introduction}

Document compilation engines are critical developer infrastructure for
science and industry: they turn markup into the PDFs that implement
publication, theses, CVs, and reports. The ecosystem is fragmented, classical
TeX engines (pdfLaTeX, XeLaTeX, LuaLaTeX), modern wrappers (Tectonic),
new languages (Typst), and Markdown pipelines (pandoc backends). Choosing an
engine affects CI time, local editor feedback, font support, and whether a
template compiles at all.

\paragraph{Build infrastructure, not only ``LaTeX engines''}
Compilation is embedded in the same class of systems as language toolchains
and package managers: continuous-integration jobs that gate pull requests,
Docker/container images that pin TeX~Live or MiKTeX for reproducible
builds, arXiv and journal submission pipelines that recompile sources under
operator-controlled engines, and campus or lab templates that must survive
years of OS upgrades. A slow or non-portable engine is a \emph{build-system}
cost (queue time, flaky CI, broken Docker layers), not a typesetting
preference.

\paragraph{Why engine reliability matters for reproducible science and CI}
Reproducible science assumes that the same markup yields a usable PDF under
the toolchain a collaborator, container, or publisher actually runs. When
engines or TeX distributions diverge, CI can pass on one host and fail on
another, arXiv recompiles can surface silent layout or font changes, and
Docker layers that pin the ``wrong'' package set become permanent flakiness.
That is an \emph{infrastructure reliability} problem: we need shared
measurement substrate (corpus, tasks, multi-OS protocol) and reusable tools
such as \textbf{ETB-Porta}, a recommender and portability gate others can
embed in CI, rather than one-off single-machine rankings.

Yet the community has \textbf{no evidence-based decision framework} and
\textbf{no shared evaluation infrastructure} for this choice. Advice lives in
blog posts and single-template anecdotes. Tan and Rigger showed that TeX
engines can emit inconsistent documents from identical correct
sources~\cite{tan2024tex}, a reliability phenomenon, but did not provide a
shared multi-platform measurement stack. Repair and generation benchmarks
typically pin a single engine~\cite{zhu2022eqfix,wang2026texocr}, so their
results cannot transfer across
toolchains~\cite{just2014defects4j,legoues2015manybugs,jimenez2024swebench,chen2021humaneval}.

\paragraph{What is missing}
Infrastructure papers that change a field (Defects4J, SWE-bench, HumanEval)
define \emph{tasks}, \emph{oracles}, \emph{protocols}, and a \emph{corpus}
others can reuse. Document compilation lacks such shared substrate. Without
it, every study invents its own seeds, timeouts, and success
definitions, results cannot accumulate.

\paragraph{This paper}
We introduce \textbf{Engine-Transfer-Bench (ETB)} as empirical
infrastructure, the benchmark \emph{is} the reusable substrate. The findings
characterize ecosystem reliability:
\begin{enumerate}
  \item A redistributable corpus of \textbf{1{,}784} working documents with
  license metadata (not synthetic mutants).
  \item Four standardized tasks with fixed metrics
  (Table~\ref{tab:tasks}).
  \item A pinned multi-engine protocol and open harness (4{,}211 native
  compiles $+$ 297 Markdown runs in our reference campaign, plus full multi-OS
  GHA replicas, Table~\ref{tab:crossos}).
  \item A \textbf{decision methodology}: condition claims on portable vs.\
  engine-specific subpopulations so selection bias cannot be mistaken for
  absolute reliability.
\end{enumerate}
Reference measurements \emph{indicate} how these engines
behave under a fixed protocol and yield principles \emph{within ETB}, not
merely a leaderboard of today's binaries.

\paragraph{Hypotheses}
\begin{description}[style=unboxed,leftmargin=0pt]
  \item[\textbf{H1} (concentrated reliability risk)]
  TeX~Live compile failures on open templates concentrate in a small
  engine-specific minority rather than distributing uniformly.
  \item[\textbf{H2} (architecture over lineage for latency)]
  On multi-engine-portable documents, latency differences are large and
  systematic (not noise), with auto-fetch/XeTeX-based Tectonic faster than
  classical TeX~Live engines under a paired design.
  \item[\textbf{H3} (residual semantic divergence)]
  Even when all engines succeed, PDF-text inconsistency
  exceeds a practical $\spdf$ threshold for a non-trivial fraction of documents
  (extending~\cite{tan2024tex}).
  \item[\textbf{H4} (architectural failure modes within ETB)]
  Dominant failures on this corpus reflect font-stack mismatch, multi-file
  layout, and asset packaging, not absence of packages on a fully provisioned
  host.
\end{description}

\paragraph{Contributions}
(1)~ETB as a reusable benchmark artifact (corpus, tasks, metrics, protocol).
(2)~A conditioned evaluation methodology (portable vs.\ engine-specific).
(3)~Hypothesis-driven findings with paired latency effect sizes and
thresholded consistency analysis.
(4)~\textbf{ETB-Porta}, a recommender$+$portability-gate system evaluated on
ETB (99.9\% primary success at mean CI cost $\approx$2.05, 78.9\% of pdf
fails caught in CI, and a hybrid$+$pdf gate with 0\% dangerous-green at cost
2.33).
(5)~A full multi-OS GHA campaign (macOS/Ubuntu/Windows, $N{=}4211$ compiles per host)
  showing Tectonic portability (gap $\le$0.9\,pp) vs.\ classic-engine
  distribution effects, plus design principles (scoped to ETB) and an adoption
  checklist.

\section{Related Work}\label{sec:related}

\paragraph{Cross-engine TeX studies}
Tan and Rigger systematically document cross-engine and cross-version
inconsistencies~\cite{tan2024tex}. ETB \emph{reuses} that insight as Task~T3
and adds standardized latency/reliability tasks, open template scale, Typst
and Markdown tracks, and a decision decomposition for practitioners and tool
builders.

\paragraph{Executable benchmarks as infrastructure}
Defects4J~\cite{just2014defects4j}, ManyBugs~\cite{legoues2015manybugs},
SWE-bench~\cite{jimenez2024swebench}, and HumanEval~\cite{chen2021humaneval}
succeed because later papers adopt their tasks and oracles. ETB aims at the
same role for document compilation toolchains.

\paragraph{Document tooling}
EqFix~\cite{zhu2022eqfix}, TexOCR~\cite{wang2026texocr}, OverleafCopilot and
PaperDebugger~\cite{wen2024overleafcopilot,hou2026paperdebugger} advance
repair, reconstruction, or writing assistance, usually under a single pinned
engine. ETB is complementary substrate: a multi-engine yardstick those systems
can report against.

\paragraph{Companion: TeXFix-Bench}
In companion work, we introduce TeXFix-Bench~\cite{texfixbench}, a benchmark
for LLM-based repair of broken document sources. Engine-Transfer-Bench
complements that work: whereas TeXFix-Bench evaluates repair agents, this work
evaluates the compilation engines that serve as the repair oracle. Together,
they provide end-to-end infrastructure for document source reliability
research.

\paragraph{Mutation testing methodology}
Controlled fault injection~\cite{jia2011mutation} motivates clean oracles. ETB
instead studies \textbf{working} documents (no mutations) so engine selection
is not confounded with repair difficulty.

\paragraph{Positioning against prior benchmarks}
Table~\ref{tab:benchpos} contrasts ETB with related evaluation artifacts.
Unlike single-engine repair oracles, ETB's unit of analysis is the
\emph{toolchain}, and its primary outcome is a conditioned decision, not a
model leaderboard.

\begin{table}[t]
\caption{How ETB relates to well-known evaluation artifacts.}
\label{tab:benchpos}
\centering
\scriptsize
\setlength{\tabcolsep}{2pt}
\begin{tabular}{@{}p{0.18\columnwidth}p{0.22\columnwidth}p{0.48\columnwidth}@{}}
\toprule
\textbf{Artifact} & \textbf{Unit} & \textbf{What later work reuses} \\
\midrule
Defects4J & Java bugs $+$ tests & Fault corpus, test oracles \\
SWE-bench & GitHub issues & Task instances, harness \\
HumanEval & Coding problems & Pass@k protocol \\
Tan\&Rigger & Correct TeX sources & Inconsistency phenomenon \\
\textbf{ETB (this)} & Working docs $+$ engines & Tasks T1--T4, protocol, portable split \\
\bottomrule
\end{tabular}
\end{table}

\section{Engine-Transfer-Bench}\label{sec:benchmark}

\begin{figure}[t]
\centering
\resizebox{\columnwidth}{!}{%
\begin{tikzpicture}[
  font=\scriptsize,
  box/.style={draw=#1!80!black,fill=#1!12,rounded corners=2.2pt,
    minimum height=1.1cm,text width=1.95cm,align=center,line width=.55pt},
  arrow/.style={-{Stealth[length=1.8mm]},line width=.65pt,draw=slate!80}
]
\node[box=accentteal] (c) {Corpus\\1{,}784 docs\\licenses};
\node[box=accentblue,right=0.38cm of c] (t) {Tasks\\T1--T4};
\node[box=accentorange,right=0.38cm of t] (p) {Protocol\\pinned engines\\60\,s, isolated};
\node[box=accentpurple,right=0.38cm of p] (a) {Analysis\\conditioned sets\\effect sizes};
\draw[arrow] (c)--(t);\draw[arrow] (t)--(p);\draw[arrow] (p)--(a);
\end{tikzpicture}}
\caption{ETB evaluation pipeline. The benchmark artifact is the corpus $+$
tasks $+$ protocol. Measurements are a reference campaign on that artifact.}
\label{fig:pipeline}
\end{figure}
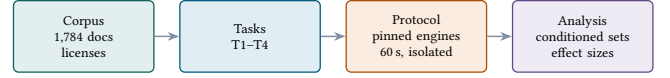

ETB is defined by four components: corpus, tasks, metrics, and protocol.
Figure~\ref{fig:pipeline} summarizes the evaluation flow.

\subsection{Corpus}

\begin{itemize}
  \item \textbf{Size:} 1{,}784 unique working documents (809 LaTeX, 975
  Typst) after license, safety, and primary-compile filters.
  \item \textbf{Sources:} GitHub Typst (604), Typst Universe (368), GitHub
  LaTeX (332), Overleaf CC gallery (275), Oleafly template-packs (107),
  CTAN examples (98).
  \item \textbf{Licenses:} MIT, CC-BY-4.0, CC0-1.0, Apache-2.0, LPPL-1.3c,
  MIT-0, each item ships attribution metadata for redistribution.
  \item \textbf{LaTeX labels:} \emph{portable/cross-engine} (702) vs.\
  \emph{engine-specific} (107), assigned at construction by multi-engine
  compile probes.
\end{itemize}

\paragraph{Why this population (bias justification)}
We deliberately exclude non-redistributable arXiv dumps and proprietary
publisher classes. ETB targets \textbf{reproducible science on open templates},
not a probability sample of all PDFs ever written. Representativeness claims
are scoped to supply-limited open-source document software, the same honesty
Defects4J applies when it scopes to projects with test suites. Construction
uses a Tectonic-oriented primary gate for LaTeX. ETB's methodology
therefore \emph{requires} conditioned analysis (portable vs.\ engine-specific)
rather than raw full-set winner-takes-all rankings.

\subsection{Tasks}

\begin{table}[t]
\caption{ETB tasks, oracles, and primary metrics.}
\label{tab:tasks}
\centering
\scriptsize
\begin{tabular}{@{}clp{0.52\columnwidth}@{}}
\toprule
\textbf{ID} & \textbf{Task} & \textbf{Oracle / metric} \\
\midrule
T1 & Reliability & Exit 0 $\wedge$ non-empty PDF within 60\,s, success rate on defined sets \\
T2 & Latency & Wall time (ms) on successful runs, paired ratios on portable set \\
T3 & Text consistency & $\spdf$ (Eq.~\ref{eq:spdf}) on portable multi-engine PDFs, flag rate at $\tau$ \\
T4 & Failure modes & Categorized root causes from logs (font, layout, assets, \ldots) \\
\bottomrule
\end{tabular}\\[1pt]
{\footnotesize Optional track T5: Markdown$\to$PDF backends (pandoc$\times$3) on a
simplicity-ranked conversion cohort ($n{=}99$ in reference campaign).}
\end{table}

\subsection{Metrics}

\paragraph{Success}
Binary: process exit code 0 and a non-empty PDF artifact.

\paragraph{Latency}
High-resolution wall clock. CPU via \code{time -p} when available. Primary
TeX latency claims use the \textbf{portable set} (all four engines succeed)
to avoid failure truncation bias.

\paragraph{PDF text similarity $\spdf$}
Let $P(\cdot)$ be the lowercase whitespace-token sequence from Poppler
\code{pdftotext}. Define token-level LCS length and Dice coefficient
\begin{equation}
\spdf(a,b)=\frac{2\,\operatorname{LCS}\!\bigl(P(a),P(b)\bigr)}{|P(a)|+|P(b)|}.
\label{eq:spdf}
\end{equation}
\textbf{Why Dice$+$LCS (not edit distance / pixels)?}
(1)~Order-sensitive token LCS captures reflow and content drops better than
bag-of-words overlap. (2)~Dice normalizes for length asymmetry. (3)~Character
edit distance is dominated by hyphenation and is $O(n^2)$ on long theses.
(4)~Visual/raster diffs are complementary future work~\cite{tan2024tex}. ETB
prioritizes a cheap, deterministic, CI-friendly oracle. Sequences $>$15k
tokens are truncated for tractability (flagged when used).

\paragraph{Failure taxonomy}
Priority rules on first-error strings: font-stack/engine-lock $\rightarrow$
\emph{font}. Aux-write/undefined control/emergency stop $\rightarrow$
\emph{structure}. Typst missing media paths $\rightarrow$ \emph{asset}, and otherwise \emph{other}. This corrects naive ``missing package'' tags when
\code{fontspec} aborts under pdfLaTeX.

\subsection{Protocol (reference campaign)}

Fresh temp directory per run (source tree read-only), 60\,s budget. Sequential
execution. Shell-escape disabled for TeX~Live. Two passes for pdf/Xe/Lua.
Tectonic and Typst use a single invocation with engine-internal re-runs. \textbf{Pinned
stack:} Tectonic 0.17.0. TeX Live 2025. Typst 0.15.1. Pandoc 3.10.1. Poppler
25.11.0. MacOS arm64 (Node harness). Zero timeouts in the reference campaign
(4{,}508 attempts, 76\,min wall).

\subsection{Conditioning sets (methodological requirement)}

\begin{itemize}
  \item \textbf{Full LaTeX set} ($N{=}809$): all seeds $\times$ four engines.
  \item \textbf{Portable set} ($N{=}702$): fair for T2/T3 (all four engines
  succeed on every document).
  \item \textbf{Engine-specific set} ($N{=}107$): reliability stress set for T1/T4.
\end{itemize}
ETB forbids interpreting Tectonic's full-set 100\% as ``compiles arbitrary
unfiltered LaTeX.''

\section{Reference Campaign Results}\label{sec:results}

We report a complete reference campaign to (i)~validate ETB's tasks and
(ii)~test H1--H4.
\emph{Note on construction bias:} the corpus passed a Tectonic-oriented
primary gate at construction (\S\ref{sec:benchmark}). Tectonic's 100\%
full-set rate in Table~\ref{tab:main} is therefore partially circular. The
non-circular results are the cross-OS campaign (Table~\ref{tab:crossos}:
Tectonic drops to 96--97\% on GHA hosts) and the conditioned engine-specific
comparison (Table~\ref{tab:portable}: Tectonic 100\% vs.\ TeX~Live 29--45\%
on documents where \emph{other} engines defined the gate). Scientific claims
below consistently state the conditioning set.

\begin{table}[t]
\caption{Full-set native outcomes (mixture of portable $+$ engine-specific).}
\label{tab:main}
\centering
\scriptsize
\setlength{\tabcolsep}{2.8pt}
\begin{tabular}{@{}lrrrrr@{}}
\toprule
\textbf{Engine} & \textbf{N} & \textbf{Succ.} & \textbf{Rate} & \textbf{Med ms} & \textbf{Mean ms} \\
\midrule
Tectonic & 809 & 809 & 100\% & 419 & 795 \\
pdfLaTeX & 809 & 733 & 90.6\% & 836 & 1{,}109 \\
XeLaTeX & 809 & 749 & 92.6\% & 1{,}135 & 1{,}472 \\
LuaLaTeX & 809 & 750 & 92.7\% & 1{,}249 & 1{,}630 \\
Typst & 975 & 928 & 95.2\% & 67 & 147 \\
\bottomrule
\end{tabular}
\end{table}

\subsection{H1: Reliability risk is concentrated}

\textbf{Supported within ETB.} On the portable set, success is 100\% for all
engines. On the engine-specific set: Tectonic 100\%,
pdfLaTeX 29\% (31/107), XeLaTeX 44\%, LuaLaTeX 45\%
(Table~\ref{tab:portable}). Across our dataset, every document that fails any
TeX~Live engine under re-measurement lies in this 107
(\code{failAnyLive}${=}107$). Full-set rates of 90--93\% are therefore a
\textbf{mixture model}: perfect success on 86.8\% of documents plus severe
loss on 13.2\%.

\begin{table}[t]
\caption{Conditioned reliability and latency (TeX). Portable $=$ H2/T2,
engine-specific $=$ H1/T1.}
\label{tab:portable}
\centering
\scriptsize
\begin{tabular}{@{}lrrrr@{}}
\toprule
 & \textbf{Tec} & \textbf{pdf} & \textbf{Xe} & \textbf{Lua} \\
\midrule
\multicolumn{5}{@{}l}{\emph{Portable} $N{=}702$ (all succeed)} \\
Success & 100\% & 100\% & 100\% & 100\% \\
Median ms & \textbf{383} & 827 & 1{,}108 & 1{,}219 \\
Mean ms & \textbf{654} & 1{,}105 & 1{,}457 & 1{,}577 \\
\midrule
\multicolumn{5}{@{}l}{\emph{Engine-specific} $N{=}107$} \\
Success & \textbf{100\%} & 29\% & 44\% & 45\% \\
\bottomrule
\end{tabular}
\end{table}

\paragraph{Effect size on reliability}
Cohen's $h$ between Tectonic (1.0) and pdfLaTeX (0.29) on the
engine-specific set is $h{\approx}2.0$ (very large). Between pdf and Xe on
that set, $h{\approx}0.31$ (small--medium): TeX~Live engines are more similar
to each other than to Tectonic on this stress slice.

\paragraph{Stratification}
Table~\ref{tab:strata} shows full-set success by category/source. Theses are
hardest for TeX~Live (pdf 70\%, Xe 74\%, Lua 72\%, $n{=}54$). GitHub/Overleaf
drive most misses. CTAN and template-packs stay near ceiling, consistent with
modern \code{fontspec} preambles in community templates vs.\ classical package
demos. This stratification is part of ETB's reporting checklist: users should
not publish a single scalar success rate without category or regime splits.

\begin{table}[t]
\caption{Full-set LaTeX success (\%) by category/source (Tec$=$100 throughout).}
\label{tab:strata}
\centering
\scriptsize
\setlength{\tabcolsep}{2.2pt}
\begin{tabular}{@{}lrrrr@{}}
\toprule
\textbf{Slice} & \textbf{N} & \textbf{pdf} & \textbf{Xe} & \textbf{Lua} \\
\midrule
package-example & 70 & 100 & 100 & 99 \\
report & 53 & 98 & 98 & 98 \\
venue & 61 & 97 & 95 & 97 \\
cv & 134 & 95 & 85 & 96 \\
beamer & 61 & 90 & 98 & 92 \\
\textbf{thesis} & 54 & \textbf{70} & \textbf{74} & \textbf{72} \\
\midrule
ctan & 98 & 100 & 100 & 99 \\
template-packs & 104 & 99 & 99 & 100 \\
overleaf & 275 & 89 & 91 & 91 \\
github-latex & 332 & 87 & 89 & 90 \\
\bottomrule
\end{tabular}
\end{table}

\subsection{H2: Latency differences are large and systematic}

\textbf{Supported} on the portable set with \textbf{paired} designs
($n{=}702$ documents, all four times observed).

\begin{table}[t]
\caption{Paired latency vs.\ Tectonic on portable set ($n{=}702$). Cliff's
$\delta{>}0$ means the other engine tends slower. Sign test: $H_0$ equal
median times.}
\label{tab:effects}
\centering
\scriptsize
\setlength{\tabcolsep}{2.2pt}
\begin{tabular}{@{}lrrrr@{}}
\toprule
\textbf{vs Tec} & \textbf{Med ratio} & \textbf{95\% boot CI} & \textbf{Cliff $\delta$} & \textbf{Sign $p$} \\
\midrule
pdfLaTeX & 2.23 & $[2.19,2.28]$ & 0.62 (large) & ${\ll}0.001$ \\
XeLaTeX & 2.91 & $[2.82,2.98]$ & 0.76 (large) & ${\ll}0.001$ \\
LuaLaTeX & 3.24 & $[3.13,3.36]$ & 0.79 (large) & ${\ll}0.001$ \\
\bottomrule
\end{tabular}\\[1pt]
{\footnotesize pdf$\rightarrow$Xe med ratio $1.34$ ($\delta{=}0.49$). Xe$\rightarrow$Lua
$1.10$ ($\delta{=}0.17$, small). pdf slower than Tec on 92\% of docs. Xe/Lua on
${>}$98\%.}
\end{table}

\textbf{Scientific reading:} under fair same-document conditions, engine
architecture (startup, font pipeline, implementation) induces
\textbf{multiplicative} latency gaps, not jitter. Tectonic's auto-fetch
XeTeX-based design is faster on the portable majority where reliability is
tied, in addition to being more reliable on exotic templates.

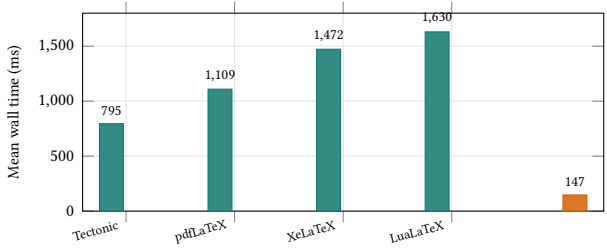
\begin{figure}[t]
\centering
\begin{tikzpicture}
\begin{axis}[
  ybar, bar width=9pt, width=\columnwidth, height=4.2cm,
  ymin=0, ymax=1800, ylabel={Mean wall time (ms)},
  symbolic x coords={Tectonic,pdfLaTeX,XeLaTeX,LuaLaTeX,Typst},
  xtick=data, x tick label style={font=\tiny,rotate=15,anchor=east},
  y tick label style={font=\scriptsize}, ylabel style={font=\scriptsize},
  grid=major, grid style={slate!20},
  nodes near coords, nodes near coords style={font=\tiny},
]
\addplot[fill=accentteal!85,draw=accentteal] coordinates {
  (Tectonic,795) (pdfLaTeX,1109) (XeLaTeX,1472) (LuaLaTeX,1630)};
\addplot[fill=accentorange!85,draw=accentorange] coordinates {(Typst,147)};
\end{axis}
\end{tikzpicture}
\caption{Full-set mean latency overview. Authoritative TeX pairwise tests:
Table~\ref{tab:effects} (portable). Typst is a separate stack (H2 non-claim).}
\label{fig:speed}
\end{figure}

\paragraph{Typst (stack-level, not same-source)}
Typst median 67\,ms on successful \code{.typ} compiles (95.2\% success). We
\textbf{do not} claim ``Typst is $11\times$ Lua on the same file''. We claim
the Typst \emph{stack} offers sub-100\,ms interactive potential that TeX
stacks do not match on this corpus.

\subsection{H3: Text inconsistency remains after multi-engine success}

\textbf{Supported within ETB.} On the portable 702, mean pairwise $\spdf$ is high
(Xe--Lua 0.982, Tectonic--pdf 0.951) but $\min\spdf{<}0.95$ flags
\textbf{25.5\%} of documents. Threshold sensitivity
(Table~\ref{tab:threshold}) shows the phenomenon is not an artifact of a
single $\tau$: 8.5\% at 0.80, 15\% at 0.90, 53\% at 0.99. Closest engines
share lineage (Xe$\leftrightarrow$Lua and Tectonic$\leftrightarrow$Xe).

\paragraph{Flag validation.}
To confirm that $\spdf$ flags reflect real content divergence rather than
\code{pdftotext} artifacts, we validated a random sample of 50 flagged pairs
(lowest-$\spdf$ engine pair per document) with an automated line-level diff
classifier, followed by author review of every sampled diff with
Unicode-normalization and whitespace-fold checks. Result: \textbf{47/50
(94\%)} show genuine content differences (dropped words, missing lines or
whole sections, divergent text), and \textbf{3/50 (6\%)} are extraction
artifacts, two Unicode NFC/NFD normalization mismatches on accented
characters and one \emph{fi}-ligature folding difference. The
$\spdf{<}0.95$ flag is thus a high-precision (94\%) indicator of real
cross-engine text divergence. The residual artifact classes suggest adding
NFC normalization and ligature folding to $\spdf$'s tokenizer in a future
revision. Per-pair adjudications ship in the artifact
(\code{spdf-validation.json}).

\begin{table}[t]
\caption{H3: inconsistency rate vs.\ threshold ($N{=}702$ portable).}
\label{tab:threshold}
\centering
\scriptsize
\begin{tabular}{@{}lrr@{}}
\toprule
$\tau$ in $\min\spdf < \tau$ & Count & Rate \\
\midrule
0.80 & 60 & 8.5\% \\
0.90 & 105 & 15.0\% \\
0.95 (default) & 179 & 25.5\% \\
0.99 & 372 & 53.0\% \\
\bottomrule
\end{tabular}
\end{table}

\begin{figure}[t]
\centering
\begin{tikzpicture}[font=\scriptsize,
  cell/.style={minimum width=1.0cm,minimum height=0.72cm,draw=white,line width=0.7pt,align=center}]
\node[cell,fill=heat6] at (0,0) {1.00};
\node[cell,fill=heat4] at (1.05,0) {0.95};
\node[cell,fill=heat5] at (2.10,0) {0.98};
\node[cell,fill=heat4] at (3.15,0) {0.96};
\node[cell,fill=heat4] at (0,-0.8) {0.95};
\node[cell,fill=heat6] at (1.05,-0.8) {1.00};
\node[cell,fill=heat5] at (2.10,-0.8) {0.97};
\node[cell,fill=heat5] at (3.15,-0.8) {0.97};
\node[cell,fill=heat5] at (0,-1.6) {0.98};
\node[cell,fill=heat5] at (1.05,-1.6) {0.97};
\node[cell,fill=heat6] at (2.10,-1.6) {1.00};
\node[cell,fill=heat6] at (3.15,-1.6) {0.98};
\node[cell,fill=heat4] at (0,-2.4) {0.96};
\node[cell,fill=heat5] at (1.05,-2.4) {0.97};
\node[cell,fill=heat6] at (2.10,-2.4) {0.98};
\node[cell,fill=heat6] at (3.15,-2.4) {1.00};
\foreach \i/\lab in {0/Tec,1/pdf,2/Xe,3/Lua} {
  \node[anchor=east] at (-0.6,-\i*0.8) {\lab};
  \node[anchor=south] at (\i*1.05,0.5) {\lab};
}
\end{tikzpicture}
\caption{Mean pairwise $\spdf$ on portable set (H3). Darker green $=$ closer.}
\label{fig:heatmap}
\end{figure}

\subsection{H4: Failures are architectural (within ETB)}

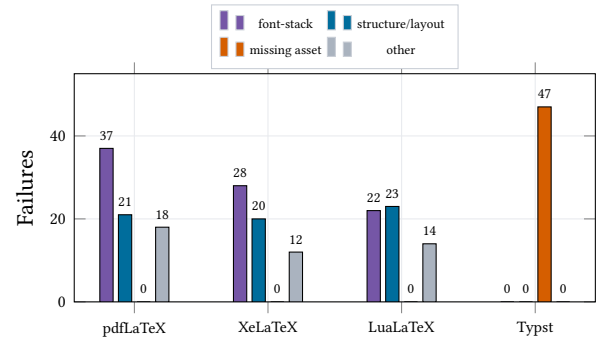
\begin{figure}[t]
\centering
\begin{tikzpicture}
\begin{axis}[
  ybar, bar width=5pt, width=\columnwidth, height=4.6cm,
  ymin=0, ymax=55, ylabel={Failures},
  symbolic x coords={pdfLaTeX,XeLaTeX,LuaLaTeX,Typst},
  xtick=data, x tick label style={font=\scriptsize},
  y tick label style={font=\scriptsize},
  legend style={font=\tiny,at={(0.5,1.02)},anchor=south,legend columns=2,draw=slate!35},
  grid=major, grid style={slate!15}, enlarge x limits=0.15,
  nodes near coords, nodes near coords style={font=\tiny},
]
\addplot[fill=accentpurple] coordinates {(pdfLaTeX,37) (XeLaTeX,28) (LuaLaTeX,22) (Typst,0)};
\addplot[fill=accentblue] coordinates {(pdfLaTeX,21) (XeLaTeX,20) (LuaLaTeX,23) (Typst,0)};
\addplot[fill=accentorange] coordinates {(pdfLaTeX,0) (XeLaTeX,0) (LuaLaTeX,0) (Typst,47)};
\addplot[fill=slate!55] coordinates {(pdfLaTeX,18) (XeLaTeX,12) (LuaLaTeX,14) (Typst,0)};
\legend{font-stack, structure/layout, missing asset, other}
\end{axis}
\end{tikzpicture}
\caption{H4: architectural failure modes within ETB (Tectonic omitted: zero failures).}
\label{fig:errors}
\end{figure}

\textbf{Supported within ETB.} Reclassified native failures
(Figure~\ref{fig:errors}): TeX~Live is dominated by \textbf{font-stack
mismatch} (pdf 37, Xe 28, Lua 22) and \textbf{multi-file/structure} issues
(pdf 21, Xe 20, Lua 23). True missing CTAN packages are $\le$1 per engine on
this provisioned host. Typst's 47 failures are \textbf{100\% missing assets}
on this corpus.

\paragraph{Root-cause lessons and representative patterns}
\begin{itemize}
  \item \textbf{Font-stack mismatch (leading TeX~Live class):}
  messages of the form ``fontspec requires XeTeX or LuaTeX'' or CTeX
  ``fontset unavailable.'' These are \emph{engine interface} failures: the
  package exists, but the PDF engine's font model is wrong. Installing more
  CTAN packages does not help.
  \item \textbf{Multi-file build layout:} ``I can't write on file
  \code{chapters/...aux}.'' Failures cluster in theses/multi-root templates
  where relative paths and working directories differ across drivers.
  \item \textbf{Package--engine interaction:} \code{tcolorbox} and similar
  packages abort with engine-specific constraints even when files resolve.
  \item \textbf{Asset packaging (Typst, 47/47):} \code{file not found
  (searched at \ldots/logo.svg)}. The language runtime is stable. The
  \emph{distribution unit} is incomplete.
\end{itemize}
ETB's T4 task is intentionally actionable: each category maps to a different
repair class (change engine, fix layout, pin fonts, ship media), a property
future automated repair systems can exploit.

\subsection{Optional track: Markdown backends}

On 99 simplicity-ranked LaTeX$\rightarrow$Markdown conversions:
pandoc$\rightarrow$HTML 100\% success (median 1.80\,s),
$\rightarrow$Tectonic 96\% (1.06\,s), $\rightarrow$Typst 85.9\% (232\,ms).
This track is \textbf{exploratory}: conversion loss confounds backend quality.
ETB includes it so future work can replace the cohort with native Markdown
without changing the task API.

\section{Principles}\label{sec:principles}

Beyond rankings, ETB yields reusable principles \emph{about this corpus and
protocol} (not universal laws of all document systems):

\begin{enumerate}
  \item \textbf{Package/auto-fetch architecture dominates reliability tails
  within ETB.}
  On engine-specific templates, Tectonic retains 100\% while TeX~Live drops to
  29--45\% despite a full TeX Live install. The gap is not ``missing
  \code{.sty} on disk'' but workflow/architecture (fetch, isolation, template
  assumptions).
  \item \textbf{Font systems dominate compatibility within ETB.}
  Font-stack mismatches are the leading TeX~Live failure class on this set.
  Engine \emph{lineage} (pdf vs.\ Xe/Lua) predicts failure better than raw
  package counts.
  \item \textbf{Runtime/implementation dominates latency on the portable
  majority (within ETB).}
  When reliability is tied, multiplicative slowdowns (median ratios
  $2.2$--$3.2\times$ vs.\ Tectonic) are the decision variable, with large
  paired effect sizes.
  \item \textbf{Success does not imply semantic identity.}
  H3 shows text-level divergence after multi-engine success on the portable
  set. Multi-engine CI needs a consistency oracle, not only exit
  codes~\cite{tan2024tex}.
  \item \textbf{Distribution packaging is part of the compiler stack.}
  Within ETB's Typst track, failures are asset packaging, not parser
  crashes, tooling must treat media as build inputs.
  \item \textbf{Conditioning is mandatory for honest evaluation.}
  Full-set averages hide a two-regime world. Benchmark users must report
  portable vs.\ engine-specific (or equivalent) splits.
\end{enumerate}

\section{ETB-Porta: Companion System}\label{sec:porta}

A benchmark becomes field infrastructure when systems are \emph{evaluated on
it} and \emph{built from its labels}. We implement \textbf{ETB-Porta}, a
companion system with two components evaluated on the ETB reference campaign:
(1)~a \textbf{static engine recommender} and (2)~a \textbf{portability gate}.
Code and full metrics ship with the artifact
(\code{harness/eval-companion.mjs}).

\subsection{Design}

\paragraph{Features (no compile required)}
From catalog metadata and the main source file we extract: font-stack signals
(\code{fontspec}, \code{unicode-math}, \verb|\setmainfont|), CJK packages,
TikZ/beamer, thesis/CV/letter categories, package count, log size, and source
stratum (GitHub, Overleaf, CTAN, templates).

\paragraph{Recommender}
Porta outputs a \emph{primary} authoring engine, an ordered \emph{authoring}
list, a \emph{CI gate} set, and per-engine risk scores. We evaluate:
\begin{itemize}
  \item \textbf{Rules:} deterministic policy distilled from ETB principles
  (font-stack $\Rightarrow$ avoid pdf-first, thesis $\Rightarrow$ multi-engine
  CI, default Tectonic$+$pdf gate).
  \item \textbf{Hybrid:} rules plus class-weighted logistic regression that
  predicts pdfLaTeX failure. High risk demotes pdf from primary and
  \emph{forces} pdf into CI for venue safety.
\end{itemize}

\paragraph{Portability gate}
Given required engines $R$ and compile outcomes,
$\mathrm{pass}\Leftrightarrow \forall e\in R:\mathrm{success}(e)$
(optional $\min\spdf\ge\tau$ in the online API). The gate operationalizes
multi-engine CI without always running all $|R|{=}4$ engines.

\subsection{Experiment}

\paragraph{Data}
All $N{=}809$ LaTeX ETB documents with complete four-engine outcomes.
pdfLaTeX failure base rate $=9.4\%$. Predictor features do not include
compile outcomes as inputs.

\paragraph{Protocol}
Five-fold out-of-fold (OOF) logistic regression for fail-pdf metrics,
class-weighted L2 training. Threshold chosen for best OOF F1. Recommenders and
gates evaluated on the full set against realized ETB outcomes. Baselines:
always-pdf, always-Tectonic, always-all-four. Gates: single-engine, all-four,
hybrid, hybrid$+$forced pdf.

\paragraph{System hypotheses}
\textbf{S-H1:} static features predict pdf failure (AUC$\gg 0.5$).
\textbf{S-H2:} Porta primary success $\ge 99\%$ at CI cost $\ll 4$.
\textbf{S-H3:} Porta gate cuts \emph{dangerous green} (CI green while pdf fails
and pdf was not scheduled) vs.\ Tectonic-only, cheaper than all-four.
\textbf{S-H4:} On engine-specific templates, always-pdf primary collapses,
Porta does not.

\subsection{Results}

\paragraph{S-H1 (supported)}
With include-graph features, OOF LogReg achieves AUC \textbf{0.89}, F1
\textbf{0.64} (precision 0.68, recall 0.61) vs.\ a font/thesis rule baseline
F1 0.36 (AUC 0.78). Failures are partially predictable \emph{before} compile.

\begin{table}[t]
\caption{Predicting pdfLaTeX failure from static$+$graph features ($N{=}809$, OOF).}
\label{tab:porta-pred}
\centering
\scriptsize
\begin{tabular}{@{}lrrrr@{}}
\toprule
\textbf{Method} & \textbf{AUC} & \textbf{Prec.} & \textbf{Rec.} & \textbf{F1} \\
\midrule
Rule (font/CJK/thesis) & 0.78 & 0.23 & 0.75 & 0.36 \\
\textbf{LogReg OOF} & \textbf{0.89} & \textbf{0.68} & 0.61 & \textbf{0.64} \\
\bottomrule
\end{tabular}
\end{table}

\paragraph{S-H2 \& S-H4 (supported)}
Table~\ref{tab:porta-rec}: always-pdfLaTeX achieves only \textbf{29\%} primary
success on engine-specific templates. Always-Tectonic is 100\% primary but
catches \textbf{0\%} of pdf failures in CI. Porta hybrid keeps
\textbf{99.9\%} primary success (99.1\% on engine-specific) at mean CI cost
\textbf{2.05} vs.\ 4.00 for all-four, while catching \textbf{78.9\%} of pdf
failures in CI.

\begin{table}[t]
\caption{Recommender comparison on ETB LaTeX set ($N{=}809$).}
\label{tab:porta-rec}
\centering
\scriptsize
\setlength{\tabcolsep}{2.2pt}
\begin{tabular}{@{}lrrrr@{}}
\toprule
\textbf{System} & \textbf{Prim.ok} & \textbf{Spec.ok} & \textbf{CI cost} & \textbf{pdf catch} \\
\midrule
always pdfLaTeX & 90.6\% & 29.0\% & 1.00 & 100\% \\
always Tectonic & 100\% & 100\% & 1.00 & 0\% \\
always all-four & 100\% & 100\% & 4.00 & 100\% \\
Porta rules & 99.9\% & 99.1\% & 2.00 & 40.8\% \\
\textbf{Porta hybrid} & \textbf{99.9\%} & \textbf{99.1\%} & \textbf{2.05} & \textbf{78.9\%} \\
\bottomrule
\end{tabular}
\end{table}

\paragraph{S-H3 (supported)}
Table~\ref{tab:porta-gate}: Tectonic-only CI passes on all documents but
yields \textbf{9.39\% dangerous green} for arXiv-style workflows
(pdf fails silently). All-four removes danger at cost 4. \textbf{hybrid$+$pdf} achieves
\textbf{0\%} dangerous green at mean cost \textbf{2.33} ($\approx$42\% fewer
CI engines than all-four). Live multi-engine$+$$\spdf$ gating is implemented
in the \code{etb-porta gate} CLI for online CI.

\begin{table}[t]
\caption{Portability gate policies. \emph{Dangerous green:} gate passes while
pdfLaTeX fails and pdf was not required.}
\label{tab:porta-gate}
\centering
\scriptsize
\begin{tabular}{@{}lrrr@{}}
\toprule
\textbf{Gate} & \textbf{Pass} & \textbf{Cost} & \textbf{Dangerous green} \\
\midrule
Tectonic only & 100\% & 1.00 & 9.39\% \\
pdfLaTeX only & 90.6\% & 1.00 & 0\% \\
All four engines & 86.8\% & 4.00 & 0\% \\
Porta hybrid CI & 89.6\% & 2.05 & 1.61\% \\
\textbf{Porta hybrid$+$pdf} & 88.0\% & \textbf{2.33} & \textbf{0\%} \\
\bottomrule
\end{tabular}
\end{table}

\paragraph{Cross-environment robustness}
A leave-one-source-out holdout (train on three corpus sources, test on the
held-out source) probes whether Porta's in-sample numbers survive
distribution shift. Porta hybrid primary success stays at \textbf{99.7\%} on
held-out GitHub ($n{=}332$. Pdf-fail base rate 13.3\%) and \textbf{100\%} on
held-out Overleaf ($n{=}275$. Base 11.3\%), CTAN ($n{=}98$), and
template-packs ($n{=}104$). The fail-pdf predictor's OOF F1 drops from 0.64
in-sample to 0.54 on the two held-out sources with nonzero base rates
(GitHub, Overleaf), a modest degradation consistent with source-specific
preamble conventions, while the safety-relevant recommender metrics are
essentially unchanged. Multi-OS T1 success rates on identical seeds
appear in \S\ref{sec:crossos} (GHA macOS/Ubuntu/Windows).

\subsection{What this shows about ETB}

Porta is not a generic ML demo: every label, split, and metric is defined by
ETB tasks (T1 outcomes, regime-aware evaluation on engine-specific
templates, and CI cost as an operationalization of T2 budgets). Without ETB's multi-engine
ledger, neither the fail-pdf predictor nor the dangerous-green metric would be
measurable. Conversely, Porta \emph{indicates} that ETB enables new
systems research beyond rankings.

\section{What ETB Enables}\label{sec:enables}

Beyond Porta, ETB unlocks: standardized engine release scores. Multi-engine
CI with $\spdf$ gates~\cite{tan2024tex}. Training data for richer recommenders,
multi-engine reporting for AI document tools~\cite{zhu2022eqfix,wen2024overleafcopilot,hou2026paperdebugger},
and differential testing seeds on the portable set.
ETB's corpus has been adopted as the seed pool for TeXFix-Bench
v0.4~\cite{texfixbench}, indicating that ETB serves as shared infrastructure
across document-compilation research.

\paragraph{Adoption checklist for papers using ETB}
Declare engine versions and OS, report T1 on the full, portable, and
engine-specific sets, report T2 with paired effect sizes on the portable
set, report T3 at multiple $\tau$, and release per-document JSONL.

\section{Closing the Closed Loop}\label{sec:extensions}

Beyond the in-corpus reference campaign we add a completed multi-OS GHA
campaign (\S\ref{sec:crossos}), external stress tests, and resource metrics.

\subsection{Cross-platform success rates}\label{sec:crossos}

We re-ran the full ETB compilation track on three GitHub Actions hosts with
the same catalog ($1{,}784$ seeds, $N{=}4{,}211$ host-tagged compiles per OS,
markdown skipped). Public harness: \url{https://github.com/prajwal-svm/etb-cross-os}.
Hosts: macOS arm64 (BasicTeX + collections), Ubuntu x64 (apt TeX~Live
subset), Windows x64 (Basic MiKTeX with \texttt{AutoInstall=1}). We compare
\textbf{success rates only}, wall times are not comparable across VMs.

\begin{table}[t]
\caption{GHA multi-OS T1 success rates (\%). Same seeds and harness. No
timing comparison. $N{=}809$ LaTeX docs per classic engine, $N{=}975$ Typst.}
\label{tab:crossos}
\centering
\scriptsize
\begin{tabular}{@{}lrrrrr@{}}
\toprule
\textbf{Engine} & \textbf{macOS} & \textbf{Ubuntu} & \textbf{Windows} & \textbf{Gap} & \textbf{All-3 OK} \\
\midrule
Tectonic  & 97.0 & 96.3 & 97.2 & \textbf{0.9} & 93.4\% \\
pdfLaTeX & 88.8 & 76.6 & 87.9 & 12.2 & 72.4\% \\
XeLaTeX & 88.8 & 74.3 & 91.3 & 17.0 & 70.5\% \\
LuaLaTeX & 93.4 & 73.8 & 90.7 & 19.6 & 70.1\% \\
Typst     & 95.2 & 83.0$^{\dagger}$ & 83.0$^{\dagger}$ & 12.2 & 83.0\% \\
\bottomrule
\end{tabular}\\[1pt]
{\footnotesize Gap $=$ max$-$min host rate (pp). All-3 OK $=$ fraction of docs
succeeding on every host. Pairwise outcome agreement: macOS--Windows 90.9\%,
Windows--Ubuntu 86.9\%, macOS--Ubuntu 86.2\%.\\
$^{\dagger}$Ubuntu and Windows Typst rates are \emph{identical} at
$809/975{=}83.0\%$ by raw count, not a copy-paste error. Both hosts fail on
the same 166 documents under a shared non-mac Typst/font environment. MacOS
GHA succeeds on those seeds (95.2\%). Shared non-mac CI gap, not a
LaTeX-distribution effect.}
\end{table}

\paragraph{Findings}
(1)~\textbf{Tectonic is OS-portable on CI:} rates 96.3--97.2\% (gap 0.9\,pp),
93.4\% of documents succeed on all three hosts.
(2)~\textbf{Classic engines track TeX distribution policy}, not OS kernel:
Ubuntu's fixed apt package set loses 12--20\,pp vs.\ macOS BasicTeX /
Windows MiKTeX auto-install. Windows and macOS GHA stay within $\sim$1--5\,pp
of each other for pdf/Xe/Lua.
(3)~\textbf{Typst:} macOS GHA matches a full desktop-style rate (95.2\%),
Ubuntu and Windows both sit at 83.0\% (same $809/975$ successes, see footnote), a shared non-mac CI gap (fonts/network), not a
LaTeX-distribution effect.
(4)~Disagreements concentrate on \texttt{missing\_package} / font errors under
thin package sets. This is a \emph{finding} about deployment environments, not
a harness bug.

\subsection{Single-engine overstatement}

On working ETB LaTeX sources, $H1(\textrm{Tectonic}){=}100\%$ but only
\textbf{86.8\%} succeed under all four engines, a \textbf{13.2\,pp}
overstatement if multi-engine portability is the estimand
(Table~\ref{tab:rescore}). A benign \code{strip\_comments} rewrite on 80 docs
drops all four engines together to 78.8\% ($\Delta{=}0$),
illustrating architecture-agnostic fragility under edit.

\begin{table}[t]
\caption{Multi-engine baseline rescoring. $\Delta=H1(\textrm{Tec})-H1(\textrm{all4})$.}
\label{tab:rescore}
\centering
\scriptsize
\begin{tabular}{@{}lrrr@{}}
\toprule
\textbf{Baseline} & \textbf{H1 Tec} & \textbf{H1 all4} & \textbf{$\Delta$} \\
\midrule
identity ($N{=}809$) & 100.0\% & 86.8\% & \textbf{13.2\,pp} \\
strip\_comments ($n{=}80$) & 78.8\% & 78.8\% & 0.0\,pp \\
\bottomrule
\end{tabular}
\end{table}

\subsection{External repositories and resource pilots}
\label{sec:pilots-main}

We report two \textbf{explicit pilot} campaigns in
Appendix~\ref{app:pilots} (not full-scale external validity):
(i)~held-out GitHub projects ($n{=}25$ attempted, $n{=}13$ evaluated, three dangerous greens all blocked by Porta hybrid$+$pdf), (ii)~resource pilot
($n{=}12$ portable docs, Tectonic warm $\approx 1.8\times$, pdfLaTeX lowest RSS). Expand or re-run before claiming production-scale external validity.

\section{Threats to Validity}\label{sec:threats}

\paragraph{Construct}
$\spdf$ is not layout equivalence. We report $\tau$-sensitivity and restrict
T3 to portable successes. Success does not imply publisher acceptance. T4 categories are
actionable approximations, not a formal fault ontology.

\paragraph{Internal}
Sequential timing. Pre-warmed Tectonic cache. Rule-based taxonomy (improved
priority rules, not double-annotated gold), T4 categories are rule-based and
author-adjudicated. A future revision should add an independent second rater
on a sample of failure logs. Paired sign tests and Cliff's
$\delta$ assume document independence. Shared packages may induce weak
dependence across seeds.

\paragraph{External}
Open templates only. Reference latency/consistency tables use a developer
macOS host. Multi-OS T1 claims use GHA macOS/Ubuntu/Windows
(Table~\ref{tab:crossos}) with host-tagged JSONL. Font and package sets differ
by distribution (apt vs.\ BasicTeX vs.\ MiKTeX). We report this as
environment variance, not a single ``OS effect.'' Source-holdout primary
success remains $\ge$99.7\% on major slices. The 107 engine-specific templates
are defined by construction probes, other partitions may move boundary
documents. Typst vs.\ TeX is stack-level. Markdown track is
conversion-confounded and exploratory.

\paragraph{Dataset evolution}
The dataset of 1{,}784 documents was collected from five open sources at a
single point in time. Document formatting practices, package ecosystems, and
engine implementations evolve continuously. The benchmark should be
periodically refreshed to remain representative. We document the collection
methodology to enable future replication with updated sources. Host-tagged
JSONL and the public cross-OS harness support longitudinal re-measurement
without redefining tasks T1--T4.

\paragraph{Conclusion}
Fixed-design census. Statistical tests strengthen within-corpus claims, not
ecological super-population inference. Selection dependence is handled by
conditioning sets and an explicit adoption protocol
(\S\ref{sec:enables}), not denied.

\paragraph{Benchmark adoption risk}
Like early Defects4J, ETB's impact depends on whether others use the tasks.
We mitigate by open data, pinned versions, JSONL schemas, and a short adoption
checklist. Community uptake is an external validity threat for the
\emph{artifact}, not for the reference measurements.

\section{Conclusion}\label{sec:conclusion}

We provide \textbf{reusable empirical infrastructure} and evidence that
compilation engines are \textbf{not interchangeable}. ETB ships an open corpus,
four tasks, a pinned multi-OS protocol, and conditioning sets that separate
reliability regimes from latency regimes. Host-tagged results show ecosystem
behavior under real CI hosts rather than a single developer machine. Within ETB,
H1--H4 indicate that font stacks drive compatibility, runtime architecture
drives portable latency, and exit-code success is insufficient for semantic
identity. A multi-OS GHA campaign (Table~\ref{tab:crossos}) shows that
\textbf{Tectonic is portable across macOS/Ubuntu/Windows} (gap $\le$0.9\,pp),
whereas classic-engine rates track package-set policy more than OS kernels.
\textbf{ETB-Porta} is infrastructure others can embed: 99.9\% primary-engine
success at mean CI cost 2.05 (vs.\ 4 for all-four), 78.9\% of pdfLaTeX
failures caught in hybrid CI, and a hybrid$+$pdf gate with 0\% dangerous-green
at cost 2.33. We release ETB, Porta, and the public multi-OS harness
(\url{https://github.com/prajwal-svm/etb-cross-os}) so evaluation, multi-engine
CI, and companion repair research~\cite{texfixbench} can share substrate
rather than folklore.

\appendix
\section{Reference Campaign Accounting}\label{app:account}
Native rows $809\times4+975=4{,}211$. Markdown $99\times3=297$. Total attempts
4{,}508. Wall 4{,}565\,s. Timeouts 0. Exact successes: Tec $809/809$, pdf
$733/809$, Xe $749/809$, Lua $750/809$, Typst $928/975$.

\paragraph{Multi-OS GHA campaign (Table~\ref{tab:crossos})}
Identical seeds and harness, $N{=}4{,}211$ compiles per host (markdown
skipped). Hosts: macOS arm64 (BasicTeX), Ubuntu x64 (apt TeX~Live subset),
Windows x64 (MiKTeX AutoInstall). Exact successes (macOS / Ubuntu /
Windows): Tec $785/779/786$. Pdf $718/620/711$. Xe $718/601/739$. Lua
$756/597/734$ of 809. Typst $928/809/809$ of 975. Pairwise outcome agreement:
macOS--Windows 90.9\%, Windows--Ubuntu 86.9\%, macOS--Ubuntu 86.2\%. Public
artifact: \url{https://github.com/prajwal-svm/etb-cross-os}.

\section{Extended Statistics}\label{app:stats}
Full-set p90/p99 (success only): Tec 1{,}583/5{,}904. Pdf 1{,}586/5{,}381. Xe
2{,}144/5{,}668. Lua 2{,}296/7{,}351. Typst 332/900\,ms. Mean PDF sizes: Tec
26\,KiB, pdf 92\,KiB, Xe 25\,KiB, Lua 49\,KiB, Typst 59\,KiB.

\section{Pilot Campaigns (Not Full-Scale)}\label{app:pilots}

\paragraph{External repositories (pilot, $n{=}13$ evaluated).}
We applied Porta live gates to held-out GitHub LaTeX projects ($n{=}25$
attempted, $n{=}13$ with a resolvable main file, 5 clone failures, 7 no main
\code{.tex}). On the evaluated set, Tectonic and pdfLaTeX each succeed on
only \textbf{23.1\%} ($3/13$), \textbf{three} dangerous greens (Tectonic OK,
pdf fail) occurred, all caught by Porta's hybrid$+$pdf gate.

\emph{Root cause of the $10/13$ failures.}
\textbf{(1)~Custom class dependency} (7/10): CV/thesis templates (awesome-cv,
moderncv, ucasthesis, SJTUThesis, zjuthesis, ElegantBook, ElegantPaper) ship
proprietary \code{.cls}/\code{.sty} files with multi-file \code{\\input}
layouts assuming project-root working directories. ETB's single-file corpus
model deliberately excludes this class.
\textbf{(2)~Font-stack mismatch} (3/10): templates require XeLaTeX-only
fonts not installed on the host. Tectonic resolves CTAN packages but cannot
install system fonts.
Zero failures stem from missing CTAN packages, the same architectural pattern
as H4 within ETB. This confirms that ETB's scope (single-file, redistributable
templates) is narrower than the full GitHub LaTeX ecosystem, not that
in-corpus metrics are inflated.

\paragraph{Resource pilot ($n{=}12$ portable docs).}
On 12 portable docs, Tectonic median warm recompile is \textbf{488\,ms} vs.\
cold 868\,ms ($\approx 1.8\times$). PdfLaTeX uses the least peak RSS
($\sim$56\,MiB) but almost no warm gain. PDF byte determinism is rare (0--25\%
across engines). Full multi-OS resource tables use the same harness flags,
expand $n$ before treating RSS/warm numbers as primary results.

\section*{Data Availability}
The ETB corpus (1{,}784 seeds with license metadata), host-tagged multi-OS
results, analysis summaries, harness, and the ETB-Porta evaluation are
archived on Zenodo (\url{https://doi.org/10.5281/zenodo.21831918}). The
cross-OS harness is public at
\url{https://github.com/prajwal-svm/etb-cross-os}.

\bibliographystyle{ACM-Reference-Format}
\bibliography{references}

\end{document}